\documentclass[11pt]{article}

\usepackage{amsmath}
\usepackage{enumerate}
\def\be{\begin{equation}}
\def\bea{\begin{eqnarray}}
\def\ee{\end{equation}}
\def\eea{\end{eqnarray}}

\def\la{\lambda}
\def\b{\bigskip}

\def\p{\partial}

\def\be{\begin{equation}}
\def\beaa{\begin{eqnarray}}
\def\ee{\end{equation}}
\def\eeaa{\end{eqnarray}}
\def \bea {\be}
\def \eea{\ee}
\def\b{\bigskip}

\def\ov{\over}

\def \bi{\bibitem}
\def \lab {\label}

\def \l {\lambda}
\def\foot{\footnote}
\def \adss {$AdS_5 \times S^5~$ }
\newcommand{\rf}[1]{(\ref{#1})}
\def \ov {\over}

\def \ha {{1 \over 2}}

\def \ci{\cite}
\def \bi {\bibitem}
\def \N  {{\cal N}}
\def\id{\protect{{1 \kern-.28em {\rm l}}}}
\def \vp {\varphi}

\def \no {\nonumber} 
\begin{document}

\overfullrule=0pt
\parskip=2pt
\parindent=12pt
\headheight=0in \headsep=0in \topmargin=0in \oddsidemargin=0in

\vspace{ -3cm}
\thispagestyle{empty}
\vspace{-1cm}

%\today

\rightline{Imperial-TP-AT-2017-08}
\rightline{   }

\begin{center}

\vspace{1cm}
{\Large\bf
%Target space
On    divergences in  non-minimal\\
\vspace{0.3cm}
$\N=4$  conformal supergravity 
}
\vspace{1.4cm}

{%O. Lunin$^{a,}$\footnote{olunin@albany.edu  }, R. Roiban$^{b,c,}$\footnote{radu@phys.psu.edu} \ and 
A.A. Tseytlin%$^{d,}$
\footnote{Also at Lebedev Institute, Moscow. tseytlin@imperial.ac.uk }}\\

\vskip 0.3cm

%{\em $^{a}$    Department of Physics, University at Albany (SUNY), Albany, NY 12222, USA }
\vskip 0.08cm

{\em  % $^{d}$
 The Blackett Laboratory, Imperial College, London SW7 2AZ, U.K.}

\vspace{.2cm}

\end{center}

{\baselineskip 11pt
\begin{abstract}
\noindent
We  review the question  of quantum consistency of $\N=4$ conformal supergravity in 4 dimensions. 
The UV divergences and   anomalies  of  the  standard (``minimal")   conformal supergravity where the  complex scalar 
$\vp$ is not coupled to the 
Weyl graviton kinetic   term  can be cancelled by coupling   this theory 
  to $\N=4$  super Yang-Mills   with gauge group of dimension 4. 
 The same  turns out to  be true   also  for   the ``non-minimal" $\N=4$ conformal supergravity  with 
the  action  (recently constructed in  arXiv:1609.09083) depending on an arbitrary  holomorphic function $f(\vp)$.
The    special   case    of  the  ``non-minimal" conformal supergravity   with $f= e^{2\vp}$    appears in the    twistor-string theory. 
We show  that  divergences and anomalies   do not depend on the   form  of the   function $f$ 
and thus can be cancelled  just  as in the ``minimal"  $f=1$ case 
 by coupling the  theory     to  four $\N=4$  vector multiplets.

% PRL    Phys Lett or J Phys Lett 

\end{abstract}
}

%\maketitle

\newpage
\def \N  {{\cal N}} \def \vp {\varphi}
\def \l {\lambda} 
\def \ba {\begin{align}}
\def \ea {\end{align}}
\def \lab   {\label}  \def \r {\rho}
\def \adst {AdS$_2 \times$S$^2$\ }
\def \la {\label}

\def \iffa {\iffalse}
\def \adstr {AdS$_3 \times$S$^3$\ }
\def \adss {AdS$_5 \times$S$^5$\ }

\def \z {\zeta}
\def \a {\alpha}\def \del {\partial}
\def \p {\phi} 
\def \ed {\end{document}}
\def \adn {AdS$_n \times$S$^n$\ }
\def \we {\wedge}
\def \P {\Phi}
\def \W  {{\cal W}}  \def \fo {{1\ov 4}}
%%%%%%%%%%%%%%%%%%%%%%%%%%%%%%%%%%%%%%

  \def \b {\beta}
  \def \te {\textstyle} 
  \def \L  {{\cal  L}}
  \def \H  {{\cal H}} \def \ha{{\textstyle {1\ov 2}}}

%\tableofcontents
%%%%%%%%%%%%%%%%%%%%%%%%%%%%%%%%%%%%%%

%\date{\today}

\setcounter{footnote}{0}
\setcounter{section}{0}

%\newpage 
%%%%%%%%%%%%%%%%%%%%%%%%%%%
%\newpage 
%\section{Introduction}
%%%%%%%%%%%%%%%%%%%%%%%%%%%%%%%

Conformal supergravities (CSGs)  are     $\N \leq 4$ supersymmetric extensions  of the    $(C_{mnkl})^2$  Weyl gravity  in 4 dimensions
\ci{1,2}. They  are   formally power-counting renormalizable  with one coupling constant. % (ignoring topological term).
The corresponding one-loop beta-functions  were  found  to be non-zero \ci{3} implying non-vanishing  conformal anomaly.
As the Weyl symmetry here is gauged, this  means  quantum inconsistency \ci{4}. 
The same   conclusion   was reached also  from  the analysis  of  chiral  $SU(4)$  R-symmetry gauge anomalies \ci{rv}, in agreement  
 with the fact that   all anomalies should  belong to the same $\N=4$ superconformal multiplet.
 % they   should be in the same  superconformal  multiplet with the Weyl anomaly. 

Remarkably,   it was observed   that 
 the   standard    
   $\N=4$  CSG theory  of  \ci{2} 
    can be made UV  finite  \ci{4,5,6}   and thus   anomaly-free   \ci{rv}
 by coupling it \ci{7}    to exactly  four  $\N=4$   super  Maxwell  multiplets (e.g., 
  to  $U(2)$  $\N=4$  super YM  theory).
  This was  shown   directly at  the one-loop order but should be true to all orders as the  beta-function 
   in $\N >1$   conformal supergravity   and conformal anomaly of SYM    may  receive  contributions  only  from the first loop 
   (as follows from formal   superspace  arguments as in  the SYM  case, see \ci{6}). 
   In the present case of $\N=4$  there  is also another reason  for one-loop exactness: 
  the  conformal  anomaly  is   tied by supersymmetry with $SU(4)$   chiral anomaly  which 
   has one-loop origin.

 In the  $SU(1,1)$ invariant   $\N=4$  CSG  of   \ci{2} the 
   4-derivative complex   scalar $\vp=\p + i \psi$  did not couple  to  Weyl  graviton  and  $SU(4)$    gauge field
   kinetic   terms. 
It was   conjectured in  \ci{4,5,6}   that there may exist  a   ``non-standard" % ``non-minimal"  
 (non $SU(1,1)$ invariant)  version   of $\N=4$ CSG. %   which  may be  UV finite by itself. 
 If one   assumes that  $\vp$  may ``non-minimally"  couple to 
 Weyl    term,  $f(\vp)(C_{mnkl})^2 + ... =  (1 + k_1 \p + k_2 \p^2 + ...) (C_{mnkl})^2 + ... $, then   there  will 
 be additional contributions to the beta-function  that may cancel  against the   ``minimal"   $\N=4$ CSG  beta-function.
 % for an appropriate choice of the   non-minimal couplings. 
  This  finiteness  conjecture  was, however, in an apparent contradiction with 
 the chiral  anomaly count  \ci{rv} as  ``non-minimal"   couplings  should not   % normally 
 contribute to the  chiral  anomaly  (see, e.g.,  \ci{8}).

  This   would   suggest  that  either (i) a ``non-minimal" theory    does not exist   as  non-minimal 
     scalar couplings are  inconsistent with 
  $\N=4$ supersymmetry,
   or (ii)  a   ``non-minimal" $\N=4$ CSG  exists    but its   UV divergences and thus 
  anomalies    are  the same  (i.e. non-vanishing) as in the  ``minimal"   theory.   
%  Until  recently,  there were  no indications   for existence of a  ``non-minimal" $\N=4$ CSG.  

   It is    the ``minimal"   $\N=4$ CSG  that appeared  (as the coefficient of the  log cutoff term) 
  in  the quantum  effective  action  of  $\N=4$ SYM   coupled  (in the standard $SU(1,1)$ covariant way  \ci{7}) 
  to the    conformal supergravity background \ci{9,10,11}  or
  in the  classical action of   the 5d $\N=8$   gauged supergravity 
  evaluated  on the solution of the  AdS$_5$    Dirichlet  boundary problem \ci{9}. 
  However,  one  reason to expect  that there should  be another  inequivalent  version of $\N=4$ CSG with 
 a  non-minimal coupling $f= e^{4 \vp} $  was  provided   \ci{6} by  dimensional reduction to 4d
   from 10d conformal supergravity    \ci{dw} (cf. eq.(4.23) in \ci{6}  and \rf{1e},\rf{3e}  below).
  
 Another  strong    indication  that a  ``non-minimal"   CSG 
     should   exist came   from the twistor-string theory \ci{12,13}  with   closed-string or singlet   gauge sector  
     describing   a  theory   with  $\N=4$ CSG     spectrum. 
     Twistor-string arguments   suggested   exponential dependence on the  scalar   and 
        the        3-point   scalar-graviton amplitudes  were  consistent with 
     $e^{2\vp}  ( C^-_{mnkl})^2 +  $c.c.  terms in the action  \ci{13}
      (see also \ci{14,am,ups}). 
      Furthermore, it was 
     conjectured in \ci{13}   that  in general  the  action of  $\N=4$ CSG 
      may contain  an arbitrary holomorphic function: 
        if $W= \vp + ... + \theta^4 C^-_{....} +...+ \theta^8 \del^4 \vp $ is a  linearized  chiral  $\N=4$ superfield strength, 
        then the action  may have  the following  structure $\int d^4x d^8 \theta\, E\,   f(W) +   c.c.   \to 
      \int d^4x \sqrt g \,  f(\vp) ( C^-_{mnkl})^2  + c.c. $.  
       It is an extra  assumption of  manifest 
       $SU(1,1) \simeq SL(2,R)$ invariance  (that includes
       constant  shifts of $\vp$) that     fixes  the function $f$ to be constant, i.e. leads to the  ``minimal" CSG.

Such    ``non-minimal"   $\N=4$ CSG   with the action 
depending on  an  arbitrary   holomorphic function 
was  indeed  constructed  recently in  \ci{15}. %   where  (the   bosonic part of) its  full non-linear   action was presented. 
 %The recent   construction   of the  supersymmetric   action  of $\N=4$   conformal supergravity \ci{15} 
 % leads to   a  resolution of these puzzles.
  % In general, the theory is  ``non-minimal",  with  coupling to $C^2$
  %   parametrized   by a single    function  $f$ of the scalar field   and is not  invariant under $SU(1,1)$;    
  %   the ``minimal"  CSG  theory  corresponds  to the special  case when 
  %$f$  is   constant  and   the $SU(1,1)$   symmetry is restored. 
  As we shall  explain below, the    puzzle    about  divergences  vs. anomalies   of such ``non-minimal" theory 
  is resolved  according to  point  (ii) above: 
  the  divergences   do not actually depend on a  particular  form of  the ``non-minimal"   function $f$, i.e. 
  are the same   as in the ``minimal" theory, in   
  agreement  with the    chiral   anomaly count \ci{rv}  as required by $\N=4$ supersymmetry. 
  Thus  there is no  ``non-minimal"  $\N=4$ CSG  theory which is  UV  finite   by itself
  but it   as in the ``minimal" case   it 
  can be made  finite and thus    consistent    by coupling it  to   four   $\N=4$  vector multiplets. 

    \
    
  %%%%%%%%%%%%%%%%%%%%%%%%%%%%%%%%%%%%%%
  %\section{Vanishing  of contribution of non-minimal couplings to one-loop divergences} 
  %%%%%%%%%%%%%%%%%%%%%%%%%%%%%%%%%%%%%%%%%%%%%%%
 % \renewcommand{\theequation}{2.\arabic{equation}} \setcounter{equation}{0}

  Let us  first  review  some  basic relations     \ci{4,5,6}. %  using  euclidean notation. 
  We shall concentrate only on  terms  involving the Weyl tensor $C_{mnkl}$, $SU(4)$ gauge field  $F^r_{mn}$  
  and  the scalar $\vp$   of $\N=4$   CSG. 
  The Lagrangian of the $\N=4$  CSG  contains the following ``minimal" terms:\foot{Note that the kinetic terms of the Weyl gravity and the $SU(4)$ gauge field have opposite   signs in the    $\N=4$ CSG action. 
  This  is consistent   with  the fact   that  integrating out the  ``matter" $\N=4$  vector multiplet  coupled to  conformal supergravity background   induces  the  $C^2$ term with positive (``asymptotically-free") sign   and the $F^2$ term with the 
  negative  (usual  ``non-asymptotically-free")  sign.}
  \be \la{1e} \te 
  \L= {2\ov \a^2} L_{\rm min}   \ , \qquad \qquad    L_{\rm min}=\vp^* D^4 \vp  +  {1 \ov 4 }  (C_{mnkl})^2 - {1 \ov 4 } ( F^r_{mn})^2    + ...\ . \ee  
  In what follows  we will suppress the internal  index $r=1, ...., 15$  on the $SU(4)$    field strength. 
  The   log UV divergent part of the effective action   is then 
  %\foot{We ignore the topological 
  %Euler number term. Its  coefficient vanishes in ``minimal"  $\N=4$ CSG \ci{5,6}.} 
  \be \la{2e}\te
  \Gamma_\infty  = - {1 \ov (4 \pi)^2 } \log \Lambda \,    \int d^4 x  \sqrt g \, b_4 \ , \ \ \ \ \ \ \ 
  b_4 = { 2} \beta L_{\rm min} \ , \ee
  where  the  beta-function coefficient   is equal to $\b=-2$  in the ``minimal"  $\N=4$ CSG. 
  For completeness, 
   let  us  recall that the conformal anomaly depends   also on  the  $a$-coefficient  of the Euler number density:
% Representing the   conformal anomaly as  
    $< T^m_m> = - a R^*R^* +  c  ( C^2_{mnkl}  -  F^2_{mn} + ...) = \b_1 R^*R^* +  \b_2  (  R^2_{mn} - {1\ov 3} R^2   -  \ha F^2_{mn} + ...) $
  %    , \ \ \ \   
   % W\equiv R^2_{mn} - {1\ov 3} R^2 = \ha (C^2_{mnkl} - R^*R^*), $   
   where 
   $\b_1=  c-a , \ \b_2 \equiv \b=  2 c $. 
   One   finds  \ci{ 4,6}    that $\b_1=  c-a$  vanishes  separately for $\N=4$ SYM and  $\N=4$   CSG  theories 
    which should  be a consequence of  their maximal $\N=4$ supersymmetry.
    The possibility of the  cancellation  of  the $\N=4$ CSG  beta-function  or $c$-anomaly 
     by coupling to  $\N=4$ SYM (four $\N=4$  vector multiplets) 
     is a non-trivial consequence of the negative sign  of $c_{_{\rm \N=4\ CSG}}$ \ci{3,4}:
    $c= c_{_{\rm \N=4\, CSG}} + 4 c_{_{\rm \N=4\,  SYM} }= -1 + 4 \times \fo =0$.
    For a discussion of  this cancellation  from   AdS$_5$ perspective \ci{9} see section 5 in \ci{16}.\foot{An   AdS$_5$ ``explanation"  of
      why the combination of $\N=4$ CSG  and  four $\N=4$   vector multiplets is anomaly-free or why $c_{_{\rm \N=4\, CSG}} = (- 2) \times 2\,  c_{_{\rm \N=4\,  SYM} }$
involves their   indirect relation to $\N=8$ 5d    supergravity: (i) the  partition functions   of   5d fields in  AdS$_5$
 and of  the corresponding 4d conformal fields  at the boundary 
  are  closely related \ci{gk,bbt,16} (by factor of -2)
   and thus  the  $a=c$  ``anomalies" of  $\N=8$ 5d   supergravity and  $\N=4$   4d CSG 
 are also  related by  factor of -2;  (ii) the $\N=8$ 5d   supergravity  may be viewed \ci{gun} as  a  product of two   ``doubletons"  -- 
 $\N=4$   vector multiplets -- and thus  their anomalies  are    related by factor for 2.} 
Similar  statement is true in 6 dimensions: conformal  a-anomaly of (2,0)   conformal supergravity is cancelled  by  coupling it to 26  (2,0) 
    tensor multiplets \ci{bt}.

  Next, let us    assume  that  the CSG action   may   contain  also some  non-minimal   scalar couplings. 
  The ones   that may contribute   to one-loop divergences  may be parametrized as  \ci{5,6}\foot{Here  we use Minkowski-metric 
   notation  and the  dual tensors are defined as 
  $F^*_{mn}= \ha \epsilon_{mnkl} F^{kl}$, \ $ \epsilon_{mnkl} \epsilon^{mnkl} = - 4!$.} 
  \ba \la{3e}
   \te L = \p\,  \del^4 \p   + {1 \ov 4 }  (C_{mnkl})^2 - {1 \ov 4 } ( F_{mn})^2   &+ 
  \p \big[  a_1   (C_{mnkl} )^2   +  b_1  C^{mnkl} C^*_{mnkl} -   a_2 (F_{mn} )^2 - b_2  F^{mn} F^{*}_{mn}  \no \\
  & +  \p^2 \big[ c_1  (C_{mnkl} )^2 - c_2 (F_{mn} )^2\big] \  , 
  \end{align} 
  where $\p$ stands   for any of the two real components  of  the complex scalar $\vp$. 
  Then it is straightforward to find that the additional  contributions of the  non-minimal terms \rf{3e} to the one-loop divergence
  in \rf{2e}  (coming from one-loop  diagrams with two  Weyl gravitons or  two 
   $SU(4)$ gauge fields  on external lines)
     are \ci{5,6}\foot{Note that  due to   4-derivative scalar  kinetic term  here one  gets  logarithmic UV divergences  from  both  the scalar  tadpole graphs 
  and   the mixed scalar-graviton and scalar-vector  loops. Compared to  similar expressions in 
   \ci{5,6}  here we have $b_1 \to i b_1, \ b_2 \to i b_2$ as   we use  Minkowski notation where $**=-1$.
  % The  factor of 2 difference  between the graviton   and gauge field  sectors is due to  kinetic term normalizations, 
 It is sufficient to  consider the    linearized expansion, 
   $(C_{mnkl})^2 \big|_{g_{mn}=\eta_{mn} + h_{mn}} = \ha (\del^2  h_{mn}^\perp)^2 + ..., $
   $ F^2_{mn}=2(\del_n V^\perp_m)^2 + ...$.
   A short-cut is just to  expand $C_{mnkl}$   and $F_{mn}$   near  background values, 
   integrate out  their fluctuations and  use that   for  $L= \p\,  \del^4 \p +  U \p^4$   one gets  in \rf{2e} 
    $\Delta b_4 =- U$. 
  } 
 \be \la{4e}
  \Delta b_4 = ( 4a_1^2 - 4b_1^2 - c_1) ( C_{mnkl})^2  -    ( 4 a_2^2 - 4 b_2^2  -  c_2) ( F_{mn})^2\ . \ee 
 %  Let us now  consider  the    specific  non-minimal couplings    which  are consistent with  supersymmetry, i.e. 
  %  which  may   actually appear    \ci{15}  in  $\N=4$ CSG   action. 
      Let us now  consider   the  particular  form of the   non-minimal couplings  that  actually   appear in the  $\N=4$ CSG action 
   constructed in \ci{15}. 
   As in  the   ``minimal"   CSG   \ci{2}   the 
    4-derivative (Weyl weight 0)   complex scalar   that parametrizes  the  $SU(1,1)/U(1)$ coset 
    may be described by a doublet $\P_\a$    of  complex 
   scalars  subject to the  $SU(1,1)$   invariant constraint  $\P^\a \P_\a \equiv  \P^*_1 \P_1 - \P^*_2 \P_2=1$    which is also invariant under the    ``non-dynamical"  $U(1)$  gauge symmetry   (with composite gauge field  coupled chirally to fermions). In the ``minimal"  
     theory   the 
    $SU(1,1)$  is  the 
    off-shell  global  symmetry  that acts only on  the  scalars.\foot{The  action of the $\N=4$    vector multiplet   coupled to conformal 
     supergravity    is invariant   under  this $SU(1,1)$  combined with a 
      duality  rotation of the   vector  field.  Once the  vector  multiplet  fields  are integrated out, 
     this  $SU(1,1)$   becomes an off-shell symmetry of the resulting induced action, i.e. of the minimal 
     CSG action \ci{10,11}.}
   Choosing the  $U(1)$   gauge  as $\P_1^* =\P_1$,   one may introduce the physical complex  scalar as 
    $\vp=\P_2/\P_1$ so that 
   \be\la{5e} 
    \P_1= (1-  |\vp|^2)^{-1/2} ,\qquad \qquad  \ \P_2=\vp\, (1-  |\vp|^2)^{-1/2} \ .  \ee
   The  $\N=4$ CSG action  of  \ci{15}  depends on an arbitrary holomorphic function  $\H(\P_1,\P_2)$  of the scalars 
      $\P_\a$  that is    homogeneous  of degree 0 in  its arguments. 
   Its presence breaks the global $SU(1,1)$ symmetry  so it is  natural to use  the  explicit parametrization in terms of $\vp$. 
    In the gauge \rf{5e}   we thus get 
   $ \H(\P_1,\P_2) \to f(\vp)$, where $f$ is a holomorphic function of the complex scalar $\vp$.  
   The  relevant terms in the  Lagrangian   of \ci{15}  that generalize
    the ``minimal" one  $L_0$ in \rf{1e}   are then given by (we ignore modification 
    of  the kinetic term of $\vp$ as its   is  not  relevant for the computation of  the one-loop  $C^2+ F^2$  divergences)
 %  We use Minkowski notation.}
   \be \la{6e}
\te    L_{\rm non-min}= {1 \ov 16 }  \big[   f(\vp) (C^-_{mnkl})^2  -    f(\vp) (F^{-}_{mn})^2  +  {\rm  c.c}  \big] +  \vp^* D^4 \vp    + ...\ , 
    \ee  
   where $C^-_{mnkl} = C_{mnkl} - i C^*_{mnkl}, \ F^-_{mn} = F_{mn} - i F^*_{mn}$. 
   For $f$=1  this reduces (up to a  total derivative) to  $L_{\rm non-min}$ in \rf{1e}.
   %When the function $\H$  (and thus  $f$)   is constant, the global $SU(1,1)$    symmetry is restored  and we get back the minimal 
 %  $\N=4$ CSG   of \ci{2}. 
   
    To find the terms that may  contribute to  $C^2+ F^2$  one-loop divergences  \rf{2e}
     it is sufficient to expand $f$  to first two orders in $\vp$, 
   \ba \la{7e}
  & f(\vp) = 1  +   k_1 \vp  + k_2 \vp^2 + O(\vp^3) \ , \ \qquad \ \ \ \ \  \vp= \p + i \psi \ ,  \\
 & \la{8e}\te 
 - {1 \ov 16}  f(\vp) (F^{-}_{mn} )^2  +  {\rm  c.c} = 
-  {1 \ov 4} [ 1 + k_1 \p  + k_2 (\p^2 - \psi^2) ]  (F_{mn})^2   -  {1 \ov 4}   (k_1 \psi   +  2  k_2  \p \psi )  F^{mn} F^*_{mn}  
 +  ...\ , 
   \end{align}
   and similarly  for the $ f(\vp) C^{mnkl} C^-_{mnkl}$ term in \rf{6e}. 
   The mixed $\p \psi $ term  can  not contribute to one-loop   divergences. % so  it was not included in \rf{3e}.
   Comparing \rf{6e},\rf{8e} to  the   non-minimal coupling ansatz in  \rf{3e}   we conclude that 
   for the real part  $\p$  of $\vp$    the constants  in \rf{3e} are
    $a_1=a_2=\fo  k_1, 
  \ b_1 =b_2=0,  \ c_1 =c_2= \fo k_2$,   while for the imaginary   part $\psi$ they are 
  $a_1=a_2= 0, 
  \ b_1 =b_2=\fo k_1, \   c_1=c_2 =-\fo k_2$.
   Summing up the contributions  of $\p$ and $\psi$ to the 
  divergent term \rf{4e}   we  find  that they cancel each other. 
  The reason why that happens 
  can be traced  back to the  holomorphicity of the  scalar   couplings in \rf{6e} dictated by the  $\N=4$ supersymmetry.\foot{Indeed,  to  explain   this cancellation  one    may  use 
  explicitly the holomorphicity of the scalar coupling:   given an (abelian)  theory   like 
  $ \vp^* \del^4 \vp   +  f(\vp)  (F^-_{mn})^2 +  c.c.$  one does not   generate $F^2_{mn} $   dependent quantum corrections 
  as propagators for both the scalar field   and the vector field strength
   involve  both conjugate components while the vertices are chiral. 
  We thank R. Roiban for  suggesting this argument.}
   
   We conclude  that the one-loop divergences  of the 
    ``non-minimal" CSG theory   do not depend   on the  function $f$ in \rf{6e} and thus are the same \rf{2e}  as in the ``minimal"  theory. This   means formally  that  only the  constant part of the function $f$ in in \rf{6e}   may be 
     deformed by renormalization.   
     The  quantum  consistency of the  theory, i.e.  the preservation of 
     the $\N=4$  superconformal gauge symmetry
      requires the cancelation of the  divergence, and that can be achieved again   by coupling  the  CSG  theory 
   (in  the usual $SU(1,1)$  covariant way \ci{7} not depending on the choice of the function $f$) 
   to  four  $\N=4$  vector multiplets \ci{4,rv,6}.
     
  This   suggests that a  twistor-string   theory that describes a coupled system of  $\N=4$ SYM and  
  ``non-minimal" 
  $\N=4$ CSG   can be  quantum-consistent      only for the  SYM gauge group  of dimension 4. 
  A world-sheet explanation  of  this  still  remains  an open problem (cf. \ci{13}).

%%%%%%%%%%%%%%%%%%%%%%%%%%%%%%%%%%%%%%%%%%%%%%%%%%%%%
\section*{Acknowledgements}
We would like to thank   % T. Adamo
 B. de Wit   for correspondence  and     R.   Roiban  for  useful discussions  and comments on the draft.
 We also acknowledge   helpful discussions with T. Adamo. 
This  work   was   supported by the ERC Advanced grant no. 290456,
 the  STFC Consolidated grant ST/L00044X/1
  and   the Russian Science Foundation grant 14-42-00047 at Lebedev Institute.

%%%%%%%%%%%%%%%%%%%%%%%%%%%%%%%%%%%%%%%%%%%%

%\newpage

%\appendix
%\section{Equations of motion and embedding into 10d supergravity}
%\renewcommand{\theequation}{A.\arabic{equation}}
%\setcounter{equation}{0}

\end{document}